# An Accurate Numerical Solution to the Kinetics of Breakable Filament Assembly

B.D. Ganapol
Department of Aerospace and Mechanical Engineering
University of Arizona

**ABSTRACT**

Proteinaceous aggregation occurs through self-assembly-- a process not entirely understood. In a recent article [**1**], an analytical theory for amyloid fibril growth via secondary rather than primary nucleation was presented. Remarkably, with only a single kinetic parameter, the authors were able to unify growth characteristics for a variety of experimental data. In essence, they seem to have uncovered the underlying allometric laws governing the evolution of filament elongation simply from two coupled non-linear ordinary differential equations (ODEs) stemming from a master equation. While this work adds significantly to our understanding of filament self-assembly, it required an approximate analytical solution representation. Here, we show that the same results are found by purely numerical means once a straightforward and reliable numerical solution to the set of ODEs has been established.

## 1. INTRODUCTION

The prediction of proteinaceous aggregation may be the key to understanding the pathology of a host of degenerative transmittable diseases [2]. Whether a result of external initiation or a symptom of infection, protein misfolding seems to be a central element in prion protein disease progression. Fortunately, how macromolecules polymerize into long chains from monomer nucleation lends itself to simulation through kinetic equations that include self-assembly and disassembly [3]. As with any physical description and especially for biophysical systems, the specification of meaningful rate constants is essential for a successful prediction of chain length distributions. More importantly however, is the discovery of underlying allometric principles governing fibrillogenesis. This requires forming appropriate combinations of rate constants and investigating their invariance to self-assembly. That was the objective of a recent article in SCIENCE [1] [hereafter referred to as **1**] where self- assembly promoting filamentous growth was theoretically investigated in an attempt to establish fundamental scaling laws. In the article, based on a master equation characterizing nucleated polymerization through conventional kinetics, the authors remark

> "The lack of analytical solutions to such master equations has, however, represented a challenge to the quest to establish general principles and laws governing filamentous growth."

The purpose of the present investigation is to demonstrate clearly that this is not the case and to indicate how the same fundamental scaling laws can be uncovered through purely numerical means. The necessary elements to do so are (1) an accurate numerical solution to a set of non-linear ODEs and (2) a straightforward dimensional analysis. While presented in the context of microscopic filamentous growth, on its own, the numerical approach represents a new way to consider a numerical solution to coupled non-linear ODEs. In particular, a highly accurate numerical solution algorithm is proposed rivaling the accuracy of any analytical solution, in particular that found in **1.**

We begin with the master equation describing filamentous growth through monomer nucleation. As with all physical investigation, the formulation of a rate equation, balancing creation and destruction of the physical elements, is a central consideration. At this point, we refer to the presentation of **1,** where the following master equation for the filament length distribution function appears:

$$\begin{aligned}\frac{\partial f(t,j)}{\partial t} &= 2k_+ m(t) f(t,j-1) - \\ &\quad -2k_+ m(t) f(t,j) - k_-(j-1) f(t,j) + \\ &\quad + 2k_- \sum_{i=j+1}^{\infty} f(t,i) + k_n m(t)^{n_c}\, \delta_{j,n_c}.\end{aligned} \tag{1}$$

The first term after the equals describes the creation of a filament of length $j$ from the nucleation of monomers at either end of a filament of length $j$-1. The second term represents the loss of a filament of length $j$ as it grows to length $j$+1 through secondary nucleation. The third term specifies the possibility of a filament of length $j$ breaking at any of its $j$-1 links. The fourth term refers to the contribution from all filaments of length greater than $j$ breaking at either end to form a filament of length $j$, and the final term approximates the source of spontaneous and homogeneous monomer nucleation to length $n_c$ seeding the growth of all subsequent filaments. $m(t)$ is the monomer concentration found from the following moment equations:

$$\begin{aligned}\frac{dP(t)}{dt} &= k_{-}\left[m(t)+(2n_c-1)P(t)\right]+k_n m(t)^{n_c}+k_m_{tot} \\ \frac{dm(t)}{dt} &= -2\left[k_{+}m(t)-n_c(n_c-1)k_{-}/2\right]P(t)-n_c k_n m(t)^{n_c}\end{aligned} \quad (2a)$$

where

$$P(t) \equiv \sum_{j=n_c}^{\infty} f(t,j)$$

is the filament number density; and if

$$M(t) = \sum_{j=n_c}^{\infty} jf(t,j),$$

is mass concentration, then

$$m(t) = m_{tot} - M(t), \quad (2b)$$

where $m_{tot}$ is the total protein concentration. Equations (2a,b) are to be solved with initial conditions

$$\begin{aligned}P(0) &= P_0 \\ M(0) &= M_0.\end{aligned} \quad (2c)$$

The main thrust of our investigation is to obtain a highly accurate and efficient solution to Eqs (2) and in so doing enable a methodology for discovering underlying scaling laws governing filament growth. Our focus will first be on the development of an accurate numerical solution thus reducing numerical uncertainty and providing a benchmark for verification of more comprehensive numerical algorithms in the future. We emphasize accuracy through simplicity in order to make high accuracy results available to numerical experts and non-experts alike.

## 2. A FINITE DIFFERENCE NUMERICAL ALGORITHM

A more convenient form of Eqs(2) emerges if one defines the moment vector

$$\boldsymbol{y}(t) \equiv \begin{bmatrix} P(t) \\ m(t) \end{bmatrix} \tag{3a}$$

leading to

$$\frac{d\boldsymbol{y}(t)}{dt} = \boldsymbol{A}\big(\boldsymbol{y}(t)\big)\boldsymbol{y}(t) + \boldsymbol{S}(t), \tag{3b}$$

where

$$\boldsymbol{A}\big(\boldsymbol{y}(t)\big) \equiv \begin{bmatrix} -k_-(2n_c-1) & -k_- + k_n m(t)^{n_c-1} \\ n_c(n_c-1)k_- & -2k_+P(t) - n_c k_n m(t)^{n_c-1} \end{bmatrix} \tag{3c}$$

and

$$\boldsymbol{S}(t) \equiv \begin{bmatrix} k_- m_{tot} \\ 0 \end{bmatrix}. \tag{3d}$$

The initial conditions then become

$$\boldsymbol{y}(0) \equiv \begin{bmatrix} P_0 \\ m_{tot} - M_0 \end{bmatrix}. \tag{3e}$$

By uniformly partitioning the time interval into intervals of $h \equiv t_{j+1} - t_j$ and integrating Eq(3b) over an interval, we find

$$\boldsymbol{y}_{j+1} - \boldsymbol{y}_j = \frac{h}{2}\Big[\boldsymbol{A}_{j+1}\big(\boldsymbol{y}_{j+1}\big)\boldsymbol{y}_{j+1} + \boldsymbol{A}_j\big(\boldsymbol{y}_j\big)\boldsymbol{y}_j\Big] + \frac{h}{2}\Big[\boldsymbol{S}_{j+1} + \boldsymbol{S}_j\Big], \tag{4a}$$

where a trapezoidal rule approximates the integrations. Hence, after matrix inversion, the solution is

$$\boldsymbol{y}_{j+1} = \left[ I - \frac{h}{2}\boldsymbol{A}_{j+1}\left(\boldsymbol{y}_{j+1}\right) \right]^{-1} \left\{ \left[ I + \frac{h}{2}\boldsymbol{A}_j\left(\boldsymbol{y}_j\right) \right]\boldsymbol{y}_j + \frac{h}{2}\left[ \boldsymbol{S}_{j+1} + \boldsymbol{S}_j \right] \right\}. \quad \text{(4b)}$$

However, because of the nonlinearity of the moments equation, Eq(4b) must be iteratively solved by lagging the matrix between times $t_j$ and $t_{j+1}$

$$\boldsymbol{A}_{j+1}\left(\boldsymbol{y}_{j+1}^{l}\right) = \boldsymbol{A}_{j+1}\left(\boldsymbol{y}_{j+1}^{l-1}\right).$$

The iterative solution therefore becomes

$$\begin{aligned} &\boldsymbol{y}_{j+1}^{0} \equiv \boldsymbol{y}_j \\ &\boldsymbol{y}_{j+1}^{l} = \left[ \boldsymbol{I} - \frac{h}{2}\boldsymbol{A}_{j+1}\left(\boldsymbol{y}_{j+1}^{l-1}\right) \right]^{-1} \left\{ \left[ \boldsymbol{I} + \frac{h}{2}\boldsymbol{A}_j \right]\boldsymbol{y}_j^{l-1} + \frac{h}{2}\left[ \boldsymbol{S}_{j+1} + \boldsymbol{S}_j \right] \right\} \end{aligned} \quad \text{(5)}$$

To this point, the development has followed along classical lines from which we know the solution is of second to Eqs(4) is order [4] and will contain an iteration error as well. Our immediate goal therefore is to use both of these errors to our advantage to attain the highest possible accuracy.

**a. Romberg convergence acceleration**

A well known, but relatively little used method to obtain high accuracy for any finite difference solution is Romberg acceleration. The concept is the same as that behind Romberg integration but now applied to a sequence of approximations of a solution to coupled ODEs having known error variation. As already noted, the true solution takes the second order form

$$\boldsymbol{y}\left(t_j\right) = \boldsymbol{y}_{j,0}\left(h\right) + \sum_{k=1}^{\infty} \boldsymbol{a}_{jk,0} h^{2k}, \quad \text{(6)}$$

where $\boldsymbol{y}_{j,0}\left(h\right)$ is the above finite difference approximation $\boldsymbol{y}_j\left(h\right)$ of Eqs(5) at iteration $l$, where $j$ refers to the time $t_j$ of the initial grid. Then, eliminating the first term of the error series by considering $\boldsymbol{y}_{j,0}\left(h\right)$ at the same time edit on a grid of half the original interval by evaluating $\boldsymbol{y}_{j,0}\left(h/2\right)$, one finds

$$\boldsymbol{y}_{j,1}(h) \equiv \left[ \frac{2^2 \boldsymbol{y}_{j,0}(h/2) - \boldsymbol{y}_{j,0}(h)}{2^2 - 1} \right],$$

as the next highest order approximation (order 4). The true flux representation for the corresponding approximation becomes

$$\boldsymbol{y}(t_j) = \boldsymbol{y}_{j,1}(h) + \sum_{k=2}^{\infty} \boldsymbol{a}_{j,k,1} h^{2k}.$$

Continuing to eliminate higher orders sequentially by continually halving the grid gives the following recurrence relation for increasingly higher order approximations:

$$\begin{aligned} \boldsymbol{y}_{j,0}(h) &\equiv \boldsymbol{y}_j(h) \\ \boldsymbol{y}_{j,m}(h) &\equiv \left[ \frac{2^{2m} \boldsymbol{y}_{j,m-1}(h/2) - \boldsymbol{y}_{j,m-1}(h)}{2^{2m} - 1} \right], \quad m = 1, 2, \ldots . \end{aligned} \tag{7}$$

resulting in the convergence acceleration (or extrapolation) of the solution to zero discretization. The solution at the original edit, which is now $t_{2^m j}$ on the grid refined $m$-times, becomes

$$\boldsymbol{y}(t_j) = \boldsymbol{y}_{j,m}(h) + \sum_{k=m+1}^{\infty} \boldsymbol{a}_{j,k,m} h^{2k}. \tag{8}$$

It should be apparent that Romberg convergence acceleration applies only to the original time edit, which must be inherited by grid refinement. One way to ensure arbitrary edits is to perform grid refinement between these edits.

Romberg acceleration requires that one generate the following sequence of finite difference approximations from Eqs(5) at each (inner) iteration $l$:

$$\boldsymbol{y}_{j,0}(h/2^m), \; m = 0, 1, 2, \ldots, \tag{9}$$

to give the Romberg sequence for the recurrence of Eq(7). The Romberg acceleration, therefore, simply rearranges the original sequence into a more efficiently converging one-- hence the name convergence acceleration. To test for convergence therefore, we have the choice of convergence of the original sequence

$$e_o \equiv \max_{k=1,2} \left| \frac{y_{j,0,k}\left(h/2^m\right) - y_{j,0,k}\left(h/2^{m-1}\right)}{y_{j,0,k}\left(h/2^m\right)} \right| < \varepsilon \tag{10a}$$

or the Romberg sequence

$$e_R \equiv \max_{k=1,2} \left| \frac{y_{j,m,k}\left(h\right) - y_{j,m-1,k}\left(h\right)}{y_{j,m,k}\left(h\right)} \right| < \varepsilon \,. \tag{10b}$$

Note that we base convergence on the worst relative error of the two moments of $\boldsymbol{y}$ at each edit.

In essence, we have redefined the meaning of a solution to be a sequence of solutions extrapolated to zero discretization error. No longer will a single discretization serve as the solution.

**b. Wynn-epsilon (*We*) convergence acceleration**

To treat the inner iteration resulting from the non-linearity, we use a Wynn-epsilon convergence algorithm [5] realizing that the iteration in Eqs(5) results in a sequence of solutions at each edit $t_j$ that presumably converges to the correct solution in the limit. In this way, we have created the sequence of approximations $\boldsymbol{y}_j^l$ during the iteration assumed to converge to the limit

$$\boldsymbol{y}_j \equiv \lim_{l\to\infty} \left\{ \boldsymbol{y}_j^l, l = 1,2,.... \right\}. \tag{11}$$

To accelerate convergence therefore, one can apply the Wynn-epsilon (*We*) convergence accelerator component-wise, which may result in a more rapidly converging sequence than the original.

The *We* accelerator takes the following form

$$\varepsilon_{-1}^{(l)} \equiv 0$$
$$\varepsilon_{0}^{(l)} \equiv P_j^l \text{ or } m_j^l \ \ , \ l = 0,\ldots,L$$
$$\varepsilon_{k+1}^{(l)} = \varepsilon_{k-1}^{(l+1)} + \left[ \varepsilon_{k}^{(l+1)} - \varepsilon_{k}^{(l)} \right]^{-1}, \ k = 0,\ldots,L \ ; \ l = 0,\ldots,L-k-1$$

for either moment sequence $P_j^l$ or $m_j^l$. The recurrence results in a tableau

$$\begin{array}{cccccc}
\varepsilon_0^{(0)} & \varepsilon_1^{(0)} & \varepsilon_2^{(0)} & \ldots & \varepsilon_{L-1}^{(0)} & \varepsilon_L^{(0)} \\
\varepsilon_0^{(1)} & \varepsilon_1^{(1)} & \varepsilon_2^{(1)} & \ldots & \varepsilon_{L-1}^{(1)} & \\
\varepsilon_0^{(2)} & \ldots & & \ldots & & \\
\ldots & & \varepsilon_2^{(L-2)} & & & \\
& \varepsilon_1^{(L-1)} & & & & \\
\varepsilon_0^{(L)} & & & & &
\end{array} \tag{12}$$

where each element of an even column estimates the limit. Convergence comes from interrogation of last term of the even columns $\varepsilon_i^{(L-i)}, \ i = 0,2,\ldots,2\left[L/2\right]$

$$e_{We} \equiv \left| \frac{\varepsilon_i^{(L-i)} - \varepsilon_i^{(L-i-2)}}{\varepsilon_i^{(L-i)}} \right| < \varepsilon, \ i = 2,\ldots,2\left[L/2\right]. \tag{13}$$

Actually, the *We* algorithm is a diagonal Padé approximant.

## 3. NUMERICAL VERIFICATION

The numerical solution of Eqs(3) is the application of *We* acceleration on the inner iterations of the nonlinear equation at each $t_j$ during the outer acceleration to a higher order accurate solution. In this section, we consider verification of the convergence acceleration algorithms thus far described as well as several additional convergence accelerations. Since the Romberg procedure requires a sequence of finite differences given by Eq(9), we can also attempt to accelerate this sequence through *We* acceleration. In addition, since the Romberg approximations are themselves a sequence, they can, in turn, be accelerated via the

*We* accelerator. Hence, the following four sequences are available for convergence interrogation:

– the original
– Romberg
– *We*
– *We* applied to Romberg.

**a. Verification by analytical solution**
If in Eq(3c), we let

$$k_+ \equiv 0$$
$$n_c = 1$$

then

$$\boldsymbol{A}\left(\boldsymbol{y}(t)\right) \equiv \begin{bmatrix} -k_- & -k_- + k_n \\ 0 & -k_n \end{bmatrix} \tag{14}$$

and Eqs(2) becomes linear. This case corresponds to no increase in filament length from nucleation. Filaments, however, are still able to break at their internal links as well as filaments of length larger than *j* can disassemble giving filaments of length *j*. However, only spontaneous growth from a single molecule is possible. For this case, the two differential equations for the moments are not only linear, but also they decouple into

$$\begin{aligned} \frac{dP(t)}{dt} &= k_-\left[m(t) + P(t)\right] + k_n m(t) + k_m_{tot} \\ \frac{dm(t)}{dt} &= -k_n m(t) \end{aligned} \tag{15a}$$

with solution

$$\begin{aligned} M(t) &= m_{tot} - m(t) = m_{tot}\left(1 - e^{-k_n t}\right) + M_0 e^{-k_n t} \\ P(t) &= m_{tot} + \left(P_0 - M_0\right)e^{-k_t} - \left(m_{tot} - M_0\right)e^{-k_n t}. \end{aligned} \tag{15b}$$

In Figs. 1a,b, for the number and mass concentration of filaments for a total protein concentration $m_{tot}$ of $10\mu M$, we assume the parameters of Table 1. The approach of the moments to each other and to their asymptotic distributions is clear from Fig. 1a. A comparison of the convergence behavior in terms of the relative error with respect to the true solution for the four sequences is shown in the error chart of Fig. 1b. The relative errors for each acceleration up to $t$ = 15h, are presented with a time offset for clarity. Each horizontal line represents an increased discretization through seven levels of grid refinement. Most remarkably, as the original sequence reaches an accuracy of about $10^{-5}$ in 7 grid refinements, the Romberg ($R$) and the $We$

**Table 1**
**Kinetic parameters for analytical solution**

$$k_{_} \equiv 10^{-5} s^{-1}$$

$$n_c = 1$$

$$M_0 = 0.21\ \mu\text{M}$$

$$P_0 = M_0 / 1380\ \text{M}$$

$$k_n = 22.8 k_{_}\ \text{M}^{-1} s^{-1}$$

$$m_{tot} = 10 \mu\text{M}.$$

accelerated Romberg ($We/R$) achieve this in 3 levels; and for the $We$ stand-alone, which overall was less impressive, in 4. This excellent performance clearly indicates the distinct advantage of acceleration. Indeed, we find near double precision machine accuracy. We next verify the algorithm including its nonlinear nature.

**b. Verification by Manufactured Solution**

A very powerful method of verifying the finite difference solution for nonlinear equations is through manufactured solutions [6]. In this method, we simply assume a solution, say $P_0(t)$ and $m_0(t)$, specify the source that gives this solution and finally run the numerical algorithm with the assumed source to see how well the assumed solution is reproduced. Here, we assume the solution to be the iterated analytical solution of **1**

$$P_0(t) = D_+ e^{\kappa t} + D_- e^{-\kappa t} - \frac{n_c k_n m_{tot}^{n_c - 1}}{2 k_+} \tag{16a}$$

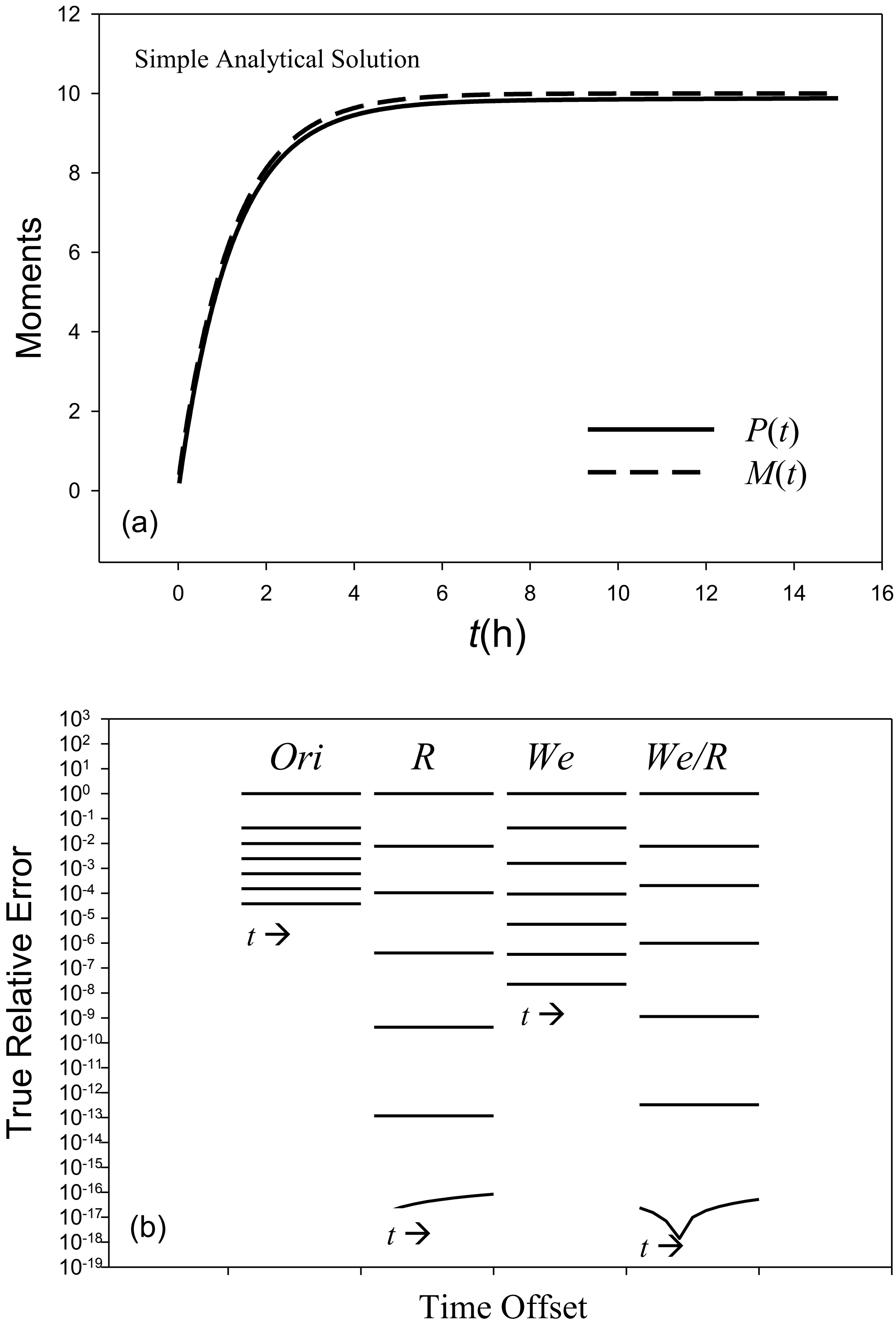

Fig. 1. (a) Exact moment traces (b) True relative error chart

$$M_0(t) = \frac{2k_+ m_{tot} D_+}{\kappa} e^{\kappa t} - \frac{2k_+ m_{tot} D_-}{\kappa} e^{-\kappa t} - \frac{k_n m_{tot}^{n_c}}{k_-}, \tag{16b}$$

where

$$\begin{aligned} \kappa &\equiv \sqrt{2m_{tot}k_+k_-} \\ D_\pm &\equiv \frac{n_c k_n m_{tot}^{n_c-1}}{4k_+} \pm \frac{k_n m_{tot}^{n_c}\kappa}{4m_{tot}k_+k_-}, \end{aligned} \tag{16c}$$

with initial conditions

$$P_0(0) = M_0(0) = 0. \tag{16d}$$

Equations (2), with known sources, become

$$\begin{aligned} \frac{dP(t)}{dt} &= -k_-\left[m(t) + (2n_c - 1)P(t)\right] + k_n m(t)^{n_c} + k_- m_{tot} + S_P(t) \\ \frac{dm(t)}{dt} &= -2\left[k_+ m(t) - n_c(n_c - 1)k_- / 2\right]P(t) - n_c k_n m(t)^{n_c} + S_m(t), \end{aligned} \tag{17a}$$

where from the assumed solution, we find

$$\begin{aligned} S_P(t) &\equiv \frac{dP_0(t)}{dt} + k_-\left[m_0(t) + (2n_c - 1)P_0(t)\right] - k_n m_0(t)^{n_c} - k_- m_{tot} \\ S_m(t) &\equiv \frac{dm_0(t)}{dt} + 2\left[k_+ m_0(t) - n_c(n_c - 1)k_- / 2\right]P_0(t) + n_c k_n m_0(t)^{n_c} \end{aligned} \tag{17b}$$

with

$$\begin{aligned} \frac{dP_0(t)}{dt} &= \kappa\left[D_+ e^{\kappa t} - D_- e^{-\kappa t}\right] \\ \frac{dm_0(t)}{dt} &= -\kappa\left[\frac{2k_+ m_{tot} D_+}{\kappa} e^{\kappa t} + \frac{2k_+ m_{tot} D_-}{\kappa} e^{-\kappa t}\right] \end{aligned} \tag{17c}$$

and

$$M_0(t) = m_{tot} - m_0(t). \tag{17d}$$

Hence, the analytical moments solutions to Eqs(17a) are

$$\begin{aligned} P(t) &= P_0(t) \\ m(t) &= m_0(t). \end{aligned}$$

We are now in position to verify the full nonlinear algorithm. For this case, we assume the parameters of Table 2. Figure 2a shows the unbounded exponential

**Table 2**
**Kinetic Parameters for the manufactured solution and converged numerical solution of Fig. 3**

$$\begin{aligned} k_+ &\equiv 5\times 10^4\,\mathrm{M^{-1}s^{-1}} \\ k_- &\equiv 2\times 10^{-8}\,\mathrm{s^{-1}} \\ n_c &= 2 \\ M_0 &= 0\mu\mathrm{M} \\ P_0 &= 0\mu\mathrm{M} \\ k_n &= 2\times 10^{-5}\,\mathrm{s^{-1}M^{-1}} \\ m_{tot} &= 5\mu\mathrm{M}. \end{aligned}$$

behavior of the moments in dynamic equilibrium characteristic of homogenous nucleation. Again, the advantage of *R* and *We/R* accelerations are evident from the error chart of Fig, 2b. The standalone *We* acceleration however is again not as effective. For this case, we have used quadruple precision arithmetic to further highlight the advantage of convergence acceleration to $10^{-30}$ relative error. The inner iteration required a maximum of 10 iterations at each *j*, and we observe nearly the same error performance as found for the analytical benchmark. In addition, when the error chart, now for 8 grid refinements, is compared to that for a simple engineering estimate of the relative error between iterations, the error estimate always conservatively overestimates the true error. The computational effort required for both this and the first benchmark was less than 5s on a Gateway 1.4 GHz laptop.

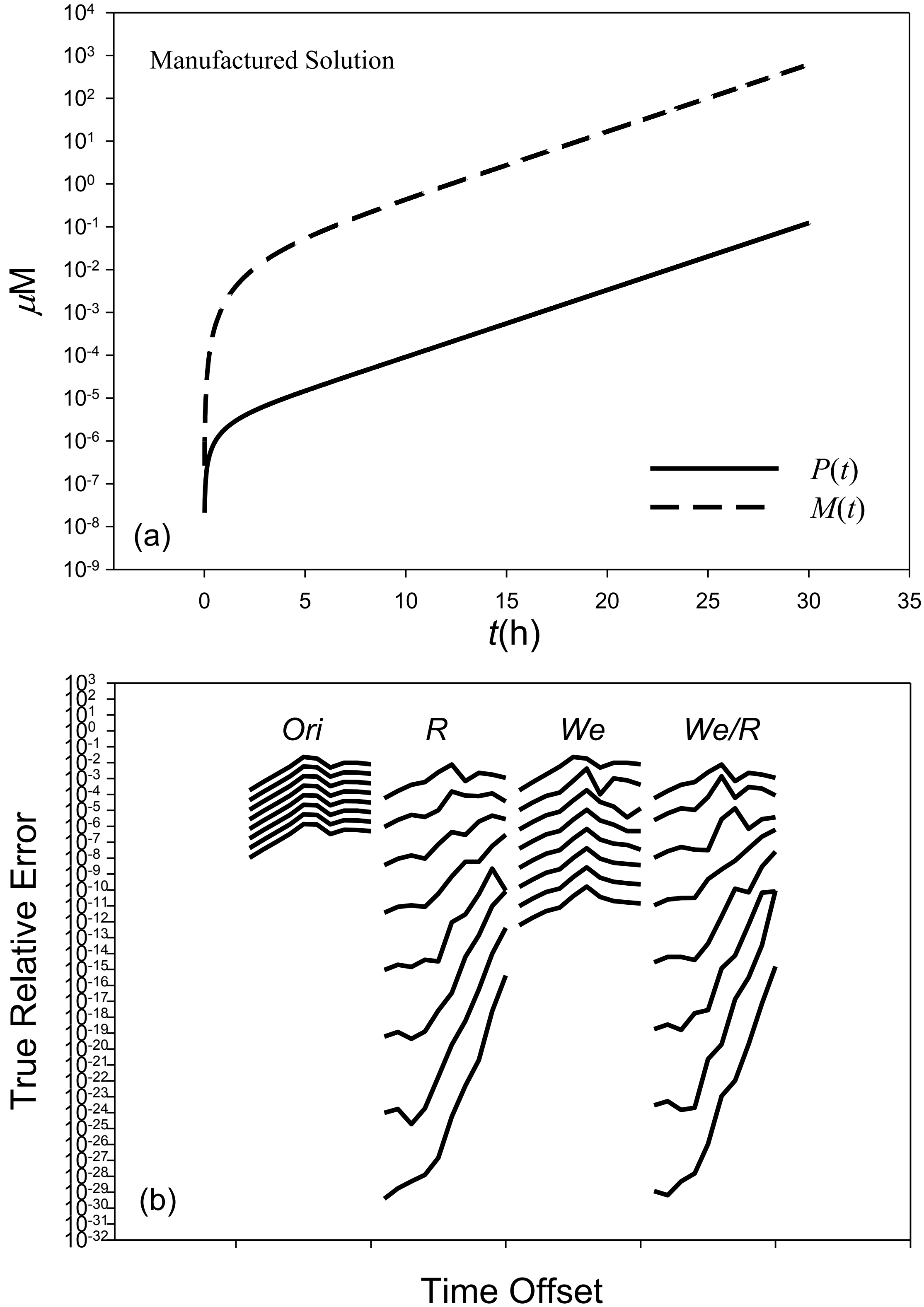

Fig. 2. (a) First iterate moment traces (b) True relative error chart

With the algorithm now verified, we can apply it to the cases found in **1** to estimate the error of their “analytical solution” derived through fixed-point iteration.

## 3. NUMERICAL CONFIRMATION

In this section, we consider the results of **1**. The solution presented there, found by analytical iteration,

$$P(t) = \frac{m_{tot}}{2n_c - 1} - \frac{m_{tot}k_-}{\kappa} e^{-(2n_c-1)k_- t} Ei\left(-C_+ e^{\kappa t}\right) + B_2 e^{-(2n_c-1)k_- t} \tag{18a}$$

$$M(t) = m_{tot}\left[1 - \exp(-C_+ e^{\kappa t} + C_- e^{-\kappa t} + \frac{k_n m_{tot}^{n_c - 1}}{k_-})\right] \tag{18b}$$

with

$$C_{\pm} \equiv \frac{k_+ P(0)}{\kappa} \pm \frac{M(0)}{2m_{tot}} \pm \frac{k_n m_{tot}^{n_c-1}}{2k_-}, \tag{18c}$$

while expressed analytically, is by no means the true analytical solution. Actually, the analytical approximation (a more appropriate label) would eventually lead to the true analytical solution if the iteration were continued indefinitely. One of several obvious flaws in this representation is that the solution for polymer mass concentration $M$ does not, in general, satisfy the initial conditions since for $t = 0$

$$M(0) = m_{tot}\left[1 - e^{-\frac{M(0)}{m_{tot}}}\right].$$

Only for a vanishingly small initial condition will this relation be satisfied. However, it is certainly an adequate approximation for an initial monomer concentration near $m_{tot}$.

Figure 3 shows the trace of the polymer concentration for the kinetic parameters of Table 2 for both the converged numerical solution and the analytical approximation of **1**. From this figure (also in **1** compared to a different numerical

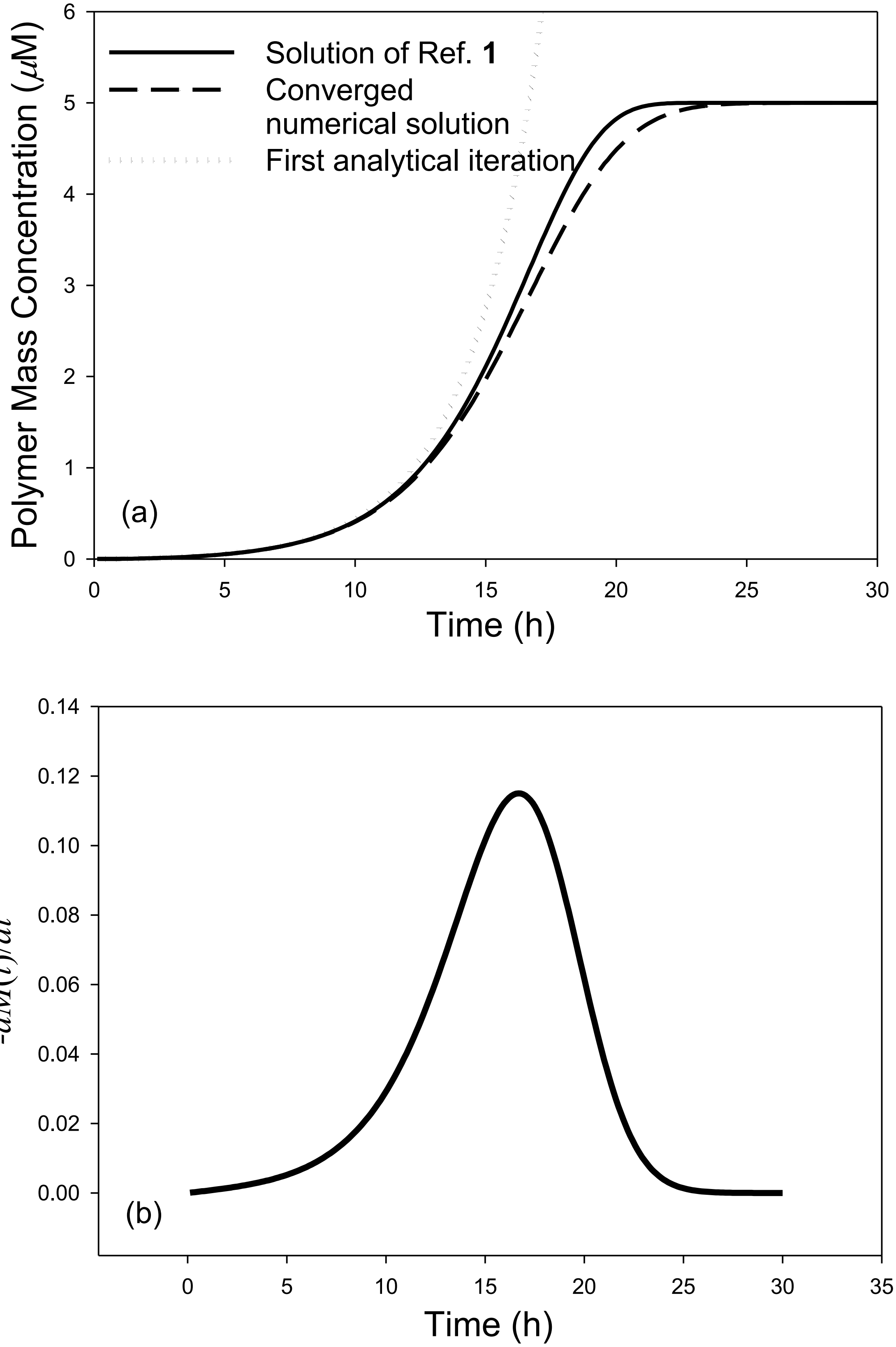

A

Fig. 3. (a) Polymer moment trace (b) Point of inflection

algorithm), the two solutions do not coincide on the steep rise of the sigmoid distribution. A closer examination however, reveals, in addition, the asymptotic values differ by a relatively small amount-- 4.999997 compared to 5.000000 for the analytical approximation. While this is physically insignificant, it is not so when comparing two supposedly highly accurate solutions, which, in principle, should agree asymptotically to at least 7 digits. What is causing this difference then?

To find the asymptotic distribution one sets the derivatives in Eqs(2) to zero giving

$$\begin{aligned} 0 &= -k_-\left[m_\infty + \left(2n_c - 1\right)P_\infty\right] + k_n m_\infty^{n_c} + k_- m_{tot} \\ 0 &= -2\left[k_+ m_\infty - n_c\left(n_c - 1\right)k_- / 2\right]P_\infty - n_c k_n m_\infty^{n_c} \end{aligned} \tag{19}$$

where $P_\infty$ and $m_\infty$ are the asymptotic values if they exist. It is a simple matter to show

$$\begin{aligned} P_\infty &= \frac{m_{tot}}{2n_c - 1} - \frac{m_\infty}{2n_c - 1} + \frac{k_n}{k_-}\frac{m_\infty^{n_c}}{2n_c - 1} \\ &2k_n k_+ m_\infty^{n_c+1} - \left[2k_+ k_- - n_c^2 k_n k_-\right]m_\infty^{n_c} - \\ &\quad - \left[n_c\left(n_c - 1\right)k_-^2 + 2k_+ k_- m_{tot}\right]m_\infty - 2k_-^2 m_{tot} = 0 \end{aligned} \tag{20}$$

For $n_c$ = 2, the later equation reduces to a cubic polynomial whose solution is known explicitly. In any case, assuming that $m_\infty << m_{tot}$, the second of Eqs(20) gives

$$m_\infty \approx \frac{k}{k_+}\frac{1}{1 + \frac{k_-}{m_{tot}k_+}} \approx \tag{21}$$

yielding

$$M_\infty \approx \quad 6\mu\text{M},$$

which agrees exactly with the numerically converged solution at $t$ = 60h and beyond. This simple exercise provides further confidence in the convergence of the proposed algorithm and highlights the inaccuracy of the analytical approximation.

One issue not addressed in **1** is the error in the lag phase $\tau_{lag}$ resulting from the difference of the analytical approximation and numerical solution at the inflection of the sigmoid response curve. $\tau_{lag}$ is a measure of the time at onset of rapid filament growth and is the time intercepted by the tangent line through the inflection point. Since the onset of rapid growth depends on this slope, the difference between the analytical approximation and the true solution should play a significant role in any analysis. An example of the inflection point, or maximum of the first derivative of the monomer population, is shown in Fig. 3b. From Fig. 4, where the phase lag for 2401 random variations of the kinetic parameters each over two decades is shown, it seems that $\tau_{lag}$ is overestimated by the analytical approximation (as expected because of its steeper slope). Fig.4 required an execution time of about 25 min for a relative error of $10^{-5}$.

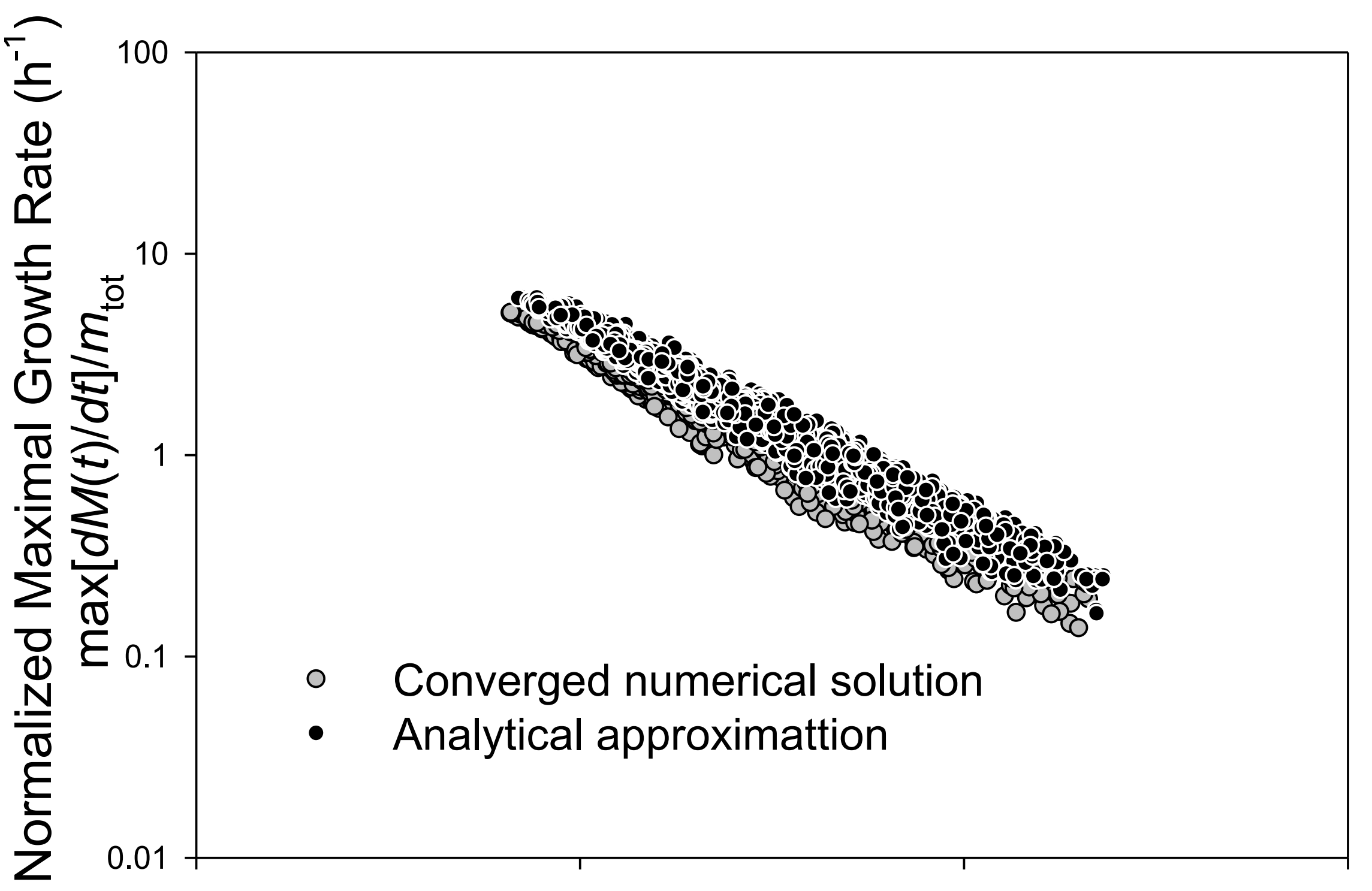

Fig. 4. Growth vs. phase lag for 2401 randomly chosen cases for the numerically converged solution and the analytical approximation

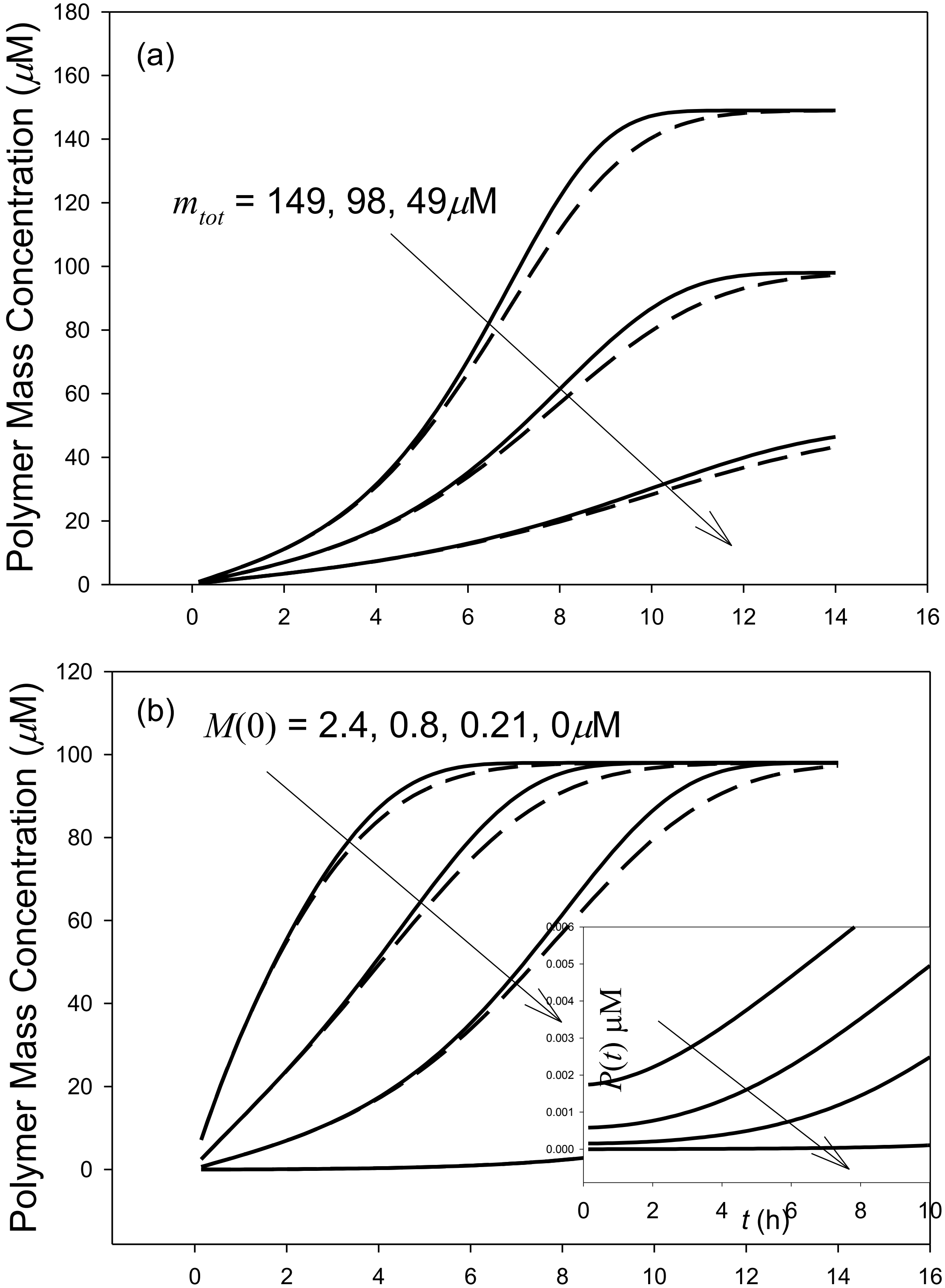

Fig. 5(a). Variation of total protein content

(b) Variation of initial conditions where $M(0)/P(0) = 1380$

Numerically converged solution

Analytical approximation

The variation between the two solutions is also apparent in comparisons with experimental measurements. The polymer concentrations shown in Fig. 5 represent such a comparison with the kinetic parameters of the analytical approximation chosen to best represent the data. In general, the numerical solution is mostly within the errors bars of the data (see **1**) and in some instances provides a more appropriate fit. The parameters for these cases are given in Tables 3a,b.

**Table 3a**
**Parameters for Figure 5a**

$$k_+ \equiv 2.9 \times 10^4 \mathrm{M^{-1}s^{-1}}$$
$$k_- \equiv 2.1 \times 10^{-9} \mathrm{s^{-1}}$$
$$n_c = 2$$
$$M(0) = 0.21 \mu\mathrm{M}$$
$$k_n / k_- = 22.8 \mathrm{M^{-1}}$$

**Table 3b**
**Parameters for Figure 5b**

$$k_+ \equiv 2.9 \times 10^4 \mathrm{M^{-1}s^{-1}}$$
$$k_- \equiv 2.1 \times 10^{-9} \mathrm{s^{-1}}$$
$$n_c = 2$$
$$M(0) / P(0) = 1380$$
$$k_n / k_- = 22.8 \mathrm{M^{-1}}$$
$$m_{tot} = 98 \mu\mathrm{M}.$$

## 4. DIMENSIONAL ANALYSIS

With the numerical algorithm verified and the numerical results of **1** confirmed, we are ready to use it in search of the scaling laws governing amyloid filament growth.

As already noted, the authors of **1** claim that to establish allometric scaling laws for proteinaceous filament growth, analytical representations (or approximations as they have developed) are required. Hence, the claim is that numerical solutions by themselves are limiting. This is certainly true if we do nothing more than just find an accurate finite difference scheme as we have done. As we shall show however, coupling a confirmed numerical solution to a dimensional analysis can potentially reveal the desired scaling laws.

With knowledge only of modeling parameters, independent and dependent variables, the solution to Eqs(2) (for $n_c = 2$) can be expressed as

$$\boldsymbol{y}(t) = \boldsymbol{F}\left[k_{-}, k_{+}, k_{n}, m_{tot}; t, P(t), M(t)\right] \tag{22}$$

If we examine the units of all quantities, one finds they are combinations of only two fundamental units-- moles (M) and time (T). In terms of these units, the listing for the parameters and variables in Eq(22) are given in Table 4. According to the Buckingham Pi theorem [7], since there are 7 parameters and variables and two fundamental units, we can express the solution in terms of five dimensionless

**Table 4**
**Dimensions of parameters and variables in Eq(22)**

$$\begin{aligned}
[k_{+}] &= \mathrm{T}^{-1} \\
[k_{-}] &= \mathrm{T}^{-1}\mathrm{M}^{-1} \\
[k_{n}] &= \mathrm{T}^{-1}\mathrm{M}^{-1} \\
[m_{tot}] &= \mathrm{M} \\
[t] &= \mathrm{T} \\
[P(t)] &= \mathrm{M} \\
[M(t)] &= \mathrm{M}
\end{aligned}$$

quantities in the form

$$\frac{\boldsymbol{y}(t)}{m_{tot}} = \boldsymbol{G}\left[\pi_i, i = 1, \ldots, 5\right] \tag{23}$$

where $\pi_i$ is a dimensionless variable. To obtain a convenient set of these variables, we begin by specifying the following cumulative dimensionless variable containing all possible choices:

$$\pi\left(a,b,\alpha,\beta,\gamma;\pi_P,\pi_M\right)=\pi\left(a,b,\alpha,\beta,\gamma\right)\pi_P\left(t\right)\pi_M\left(t\right), \qquad (24a)$$

where

$$\pi\left(a,b,\alpha,\beta,\gamma\right)\equiv k_-^a k_+^b k_n^\alpha m_{tot}^\beta t^\gamma . \qquad (24b)$$

Note that we anticipated two of the dimensionless variables as

$$\pi_P\left(t\right)\equiv\frac{P\left(t\right)}{m_{tot}},\quad \pi_M\left(t\right)\equiv\frac{M\left(t\right)}{m_{tot}}. \qquad (24c)$$

We must now determine the remaining three groups from the dimensional equivalence

$$\left[\pi\left(a,b,\alpha,\beta,\gamma\right)\right]=\mathrm{T}^{\gamma-(a+b+\alpha)}\mathrm{M}^{\beta-b-\alpha}. \qquad (25)$$

Therefore, for $\pi\left(a,b,\alpha,\beta,\gamma\right)$ to be a dimensionless quantity

$$\begin{aligned}&\gamma-\left(a+b+\alpha\right)=0\\&\beta-b-\alpha=0\end{aligned}$$

or expressing two of the five unknowns in terms of the remaining three gives

$$\begin{aligned}&a=\gamma-\beta\\&b=\beta-\alpha;\end{aligned}$$

and Eq(24b) becomes

$$\pi\left(a,b,\alpha,\beta,\gamma\right)\rightarrow\pi\left(\alpha,\beta,\gamma\right)=k_-^{\gamma-\beta}k_+^{\beta-\alpha}k_n^\alpha m_{tot}^\beta t^\gamma . \qquad (26)$$

Since the specification of the dimensionless groups in not unique, it is convenient to specify all possibilities of a given class and choose an appropriate set. For this reason, we consider all dimensionless variables to powers of 1 or -1 by letting $\alpha$, $\beta$, and $\gamma$ cycle through 0 and 1. This produces two sets of results as shown in Table 5. The choice of which three of the six groups is primarily one of convenience.

If additional values of $\alpha$, $\beta$, and $\gamma$ are assigned, additional groups are formed as products of the groups already found. In particular, for the combination (0,1,2), we find

$$\pi_7(t) = k_k_+ m_{tot} t^2 = \pi_1^2 \pi_2 = \kappa / 2, \tag{27}$$

which plays an important role in the analysis of **1** and is called the rate of multiplication of the filament population.

**Table 5**
**Dimensionless variables of powers of either 1 or -1**

| $i$ | $\alpha,\beta,\gamma$ | $\pi_i$ |
|---|---|---|
| 1 | 0,0,1 | $k_t$ |
| 2 | 0,1,0 | $m_{tot}\dfrac{k_+}{k_}$ |
| 3 | 1,0,0 | $\dfrac{k_n}{k_+}$ |
| 4 | 0,1,1 | $k_+ t$ |
| 5 | 1,1,0 | $m_{tot}\dfrac{k_n}{k_}$ |
| 6 | 1,1,1 | $k_n m_{tot} t$ |

The truth of the dimensional analysis is immediately obvious from the analytical approximation of **1** since, in terms of the dimensionless groups, Eqs(18) become

$$\begin{aligned}\pi_P(t) &= \frac{1}{3} - \frac{1}{\sqrt{2\pi_2}} e^{-3\pi_1(t)} Ei\left(-C_+ e^{\pi_1(t)\sqrt{2\pi_2}}\right) + B_2 e^{-3\pi_1(t)} \\ \pi_M(t) &= 1 - \exp(-C_+ e^{\pi_1(t)\sqrt{2\pi_2}} + C_e^{-\pi_1(t)\sqrt{2\pi_2}} + \pi_2\pi_3)\end{aligned} \tag{28a}$$

with

$$
\begin{aligned}
&C_{\pm}\left(\pi_P(0),\pi_M(0),\pi_2,\pi_3\right) \equiv \pi_P(0)\sqrt{2\pi_2} \pm \frac{1}{2}\pi_M(0) \pm \frac{1}{2}\pi_2\pi_3 \\
&B_2\left(\pi_P(0),\pi_M(0),\pi_2,\pi_3\right) = \pi_P(0) - \frac{1}{3} + \frac{1}{\sqrt{2\pi_2}}Ei\left(-C_+\right).
\end{aligned}
\qquad (28b)
$$

With the dimensionless groups as our guide, we now search for invariances in experimental data. For example, say we have the data shown in Fig. 6 [8]. We first consider $\pi_5$ and $\pi_7\left(\tau_{lag}\right)^{1/2}$ since these groupings contain all four parameters and time. Figure 7a is a plot containing the 2401 random values of the kinetic parameters used to generate Fig. 4 and these two groups. Remarkably, as predicted in **1**, we find an almost perfect linear correlation

$$
\ln \pi_5 = -1.132 - 1.414\sqrt{\pi_7\left(\tau_{lag}\right)}
$$

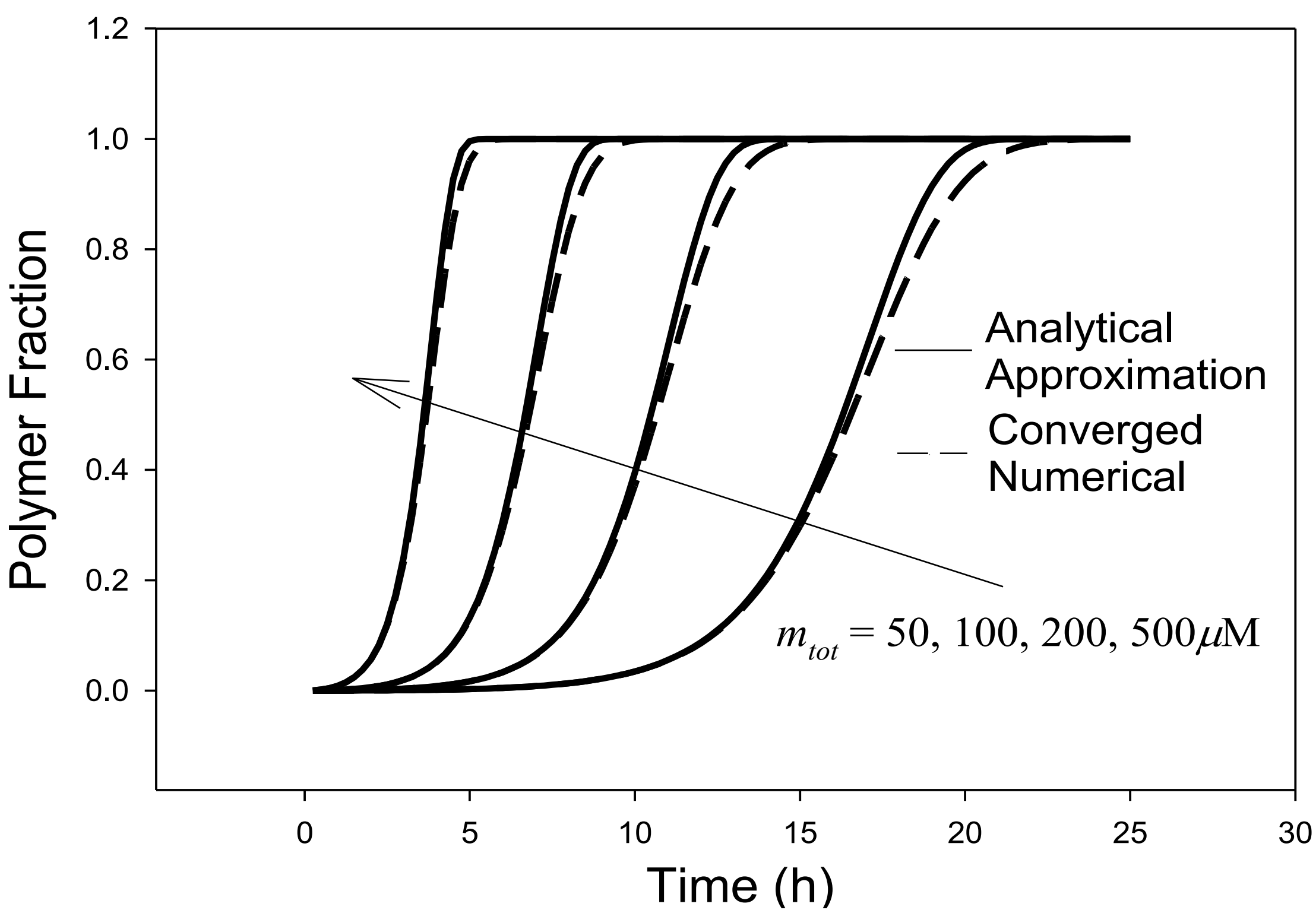

Fig. 6 Approximations to Ferguson data [8].

which when manipulated gives

$$\tau_{lag} = \frac{1}{\sqrt{2k_+ k_- m_{tot}}}\left[\ln\left(1/C_+\right) - 1.825\right] \quad (29)$$

and is the scaling found in **1** only differing in the subtracted constant (1.825 vs. 1.718) to give a longer lag time. Since this expression comes from an accurate numerical solution, Eq(29) represents a more accurate scaling law and clearly demonstrates that a purely numerical solution can indeed provide such. Finally, when the experimental data, numerically modeled in Fig 6, is added in addition to the data for Fig. 3, the correlation is confirmed.

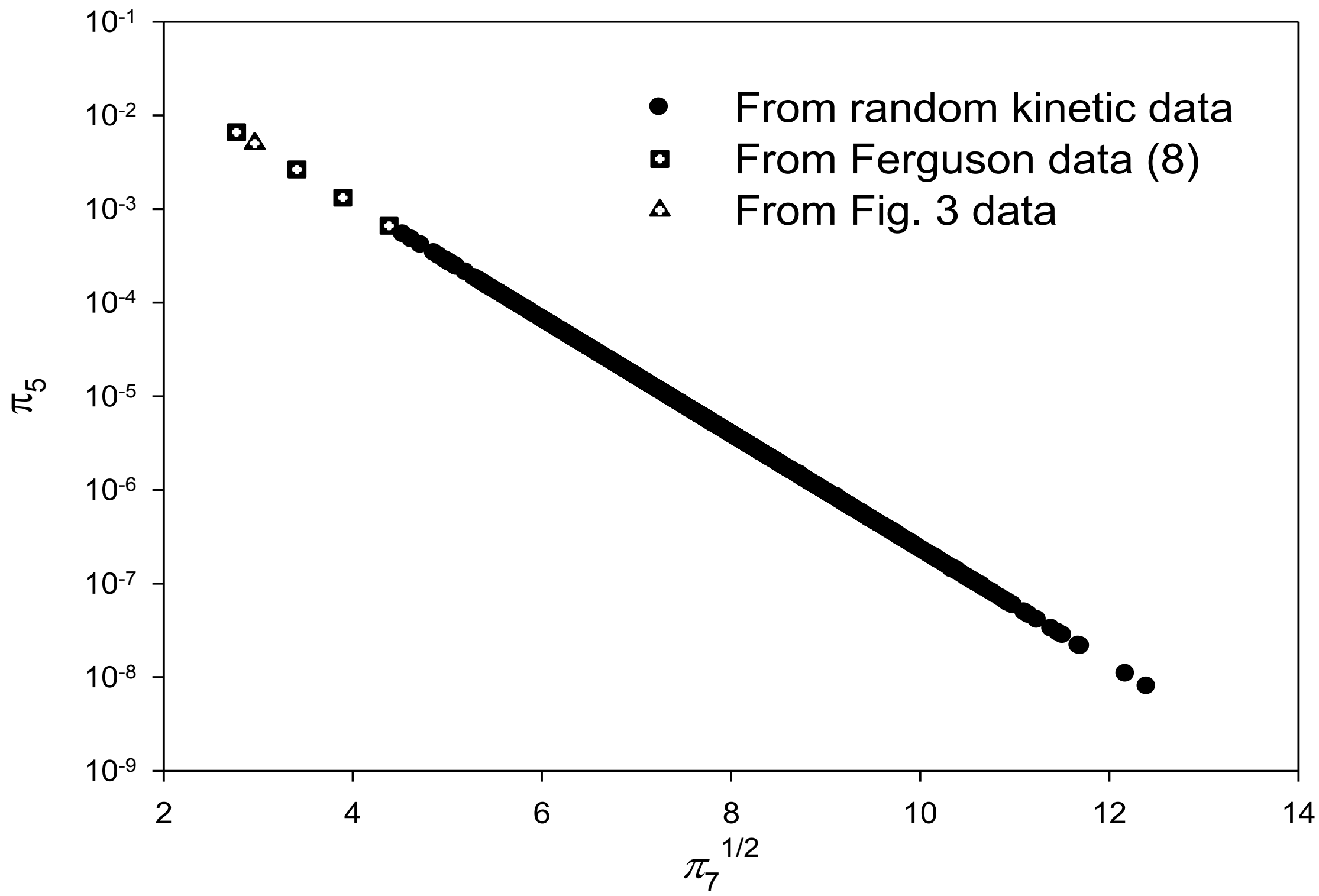

Fig. 7a. Scaling law for lag phase.

There is additional information to be found from our numerical investigation however. The normalized maximal growth $v_m$ (defined as the derivative at the inflection point normalized to the total protein content) when multiplied by the lag phase also forms a dimensionless variable. When plotted against $\pi_7\left(\tau_{lag}\right)^{1/2}$ as shown in Fig. 7b, again a nearly perfect correlation results giving the scaling law

$$\mathrm{v}_m = 0.3182\sqrt{2k_+k_-m_{tot}} + \frac{5.548\times10^{-4}}{\tau_{lag}}, \qquad (30)$$

which is again confirmed by experimental data. Hence, once the lag phase is found from Eq(29), we also find the maximal growth from Eq(30). Finally, when Eq(29) is introduced into Eq(3), $\mathrm{v}_m$ is observed to be proportional to $\kappa$, which is consistent with **1**.

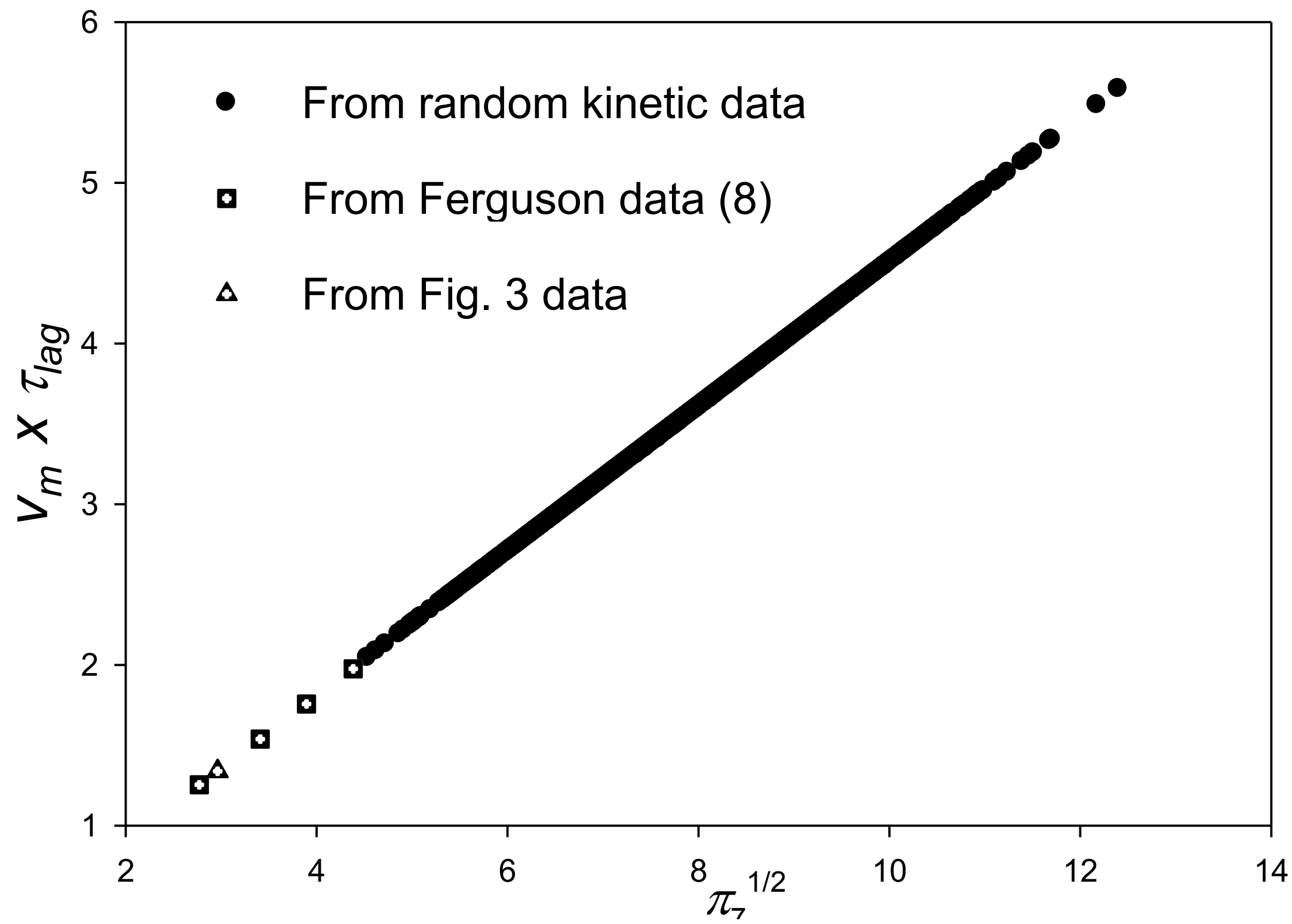

Fig. 7a Scaling law for maximal growth.

**5. FINAL COMMENTS**

We have developed a new numerical solution for the moments equations of amyloid fibril growth resulting from a master equation. The solution features high accuracy through convergence acceleration using both Romberg and Wynn-epsilon procedures to extrapolate a simple finite difference approximation to nearly zero discretization error. The numerical strategy is verified by comparison to an analytical solution of a related linear set of ODEs and to a manufactured solution based on the first iteration of the analytical approximation found in **1**. While the numerical algorithm in itself represents a new approach to solving ODEs, its value is further enhanced when coupled to a dimensional analysis. In this way, a dimensional analysis provides a guide to the underlying evolution of fibril growth from which the fundamental scaling laws emerge. Indeed, we confirm the scaling laws found in **1** to a high level of confidence since they result from a more accurate numerical solution to the governing moments equations. As a final example of this, another scaling law emerges when comparing dimensionless groups $\pi_1$ and $\pi_2$ as shown in Fig. 8. This scaling is not as strong as the two above however. Also indicated are the experimental data of Fig, 6 and the analytical data of Fig.3. While not in complete agreement with the correlation, the law is indicative of the trend.

The significance of this work is the demonstration that an analytical solution is not required to discover scaling laws. All that is needed is a reliable numerical solution and a dimensional analysis. Most likely this will be the preferred (probably the only) approach when reaction rates are dependent on the component concentrations which seems to be the direction investigations are headed. In addition, it will be the approach to the solution of the master equation itself, which is now relatively easily within reach using convergence acceleration.

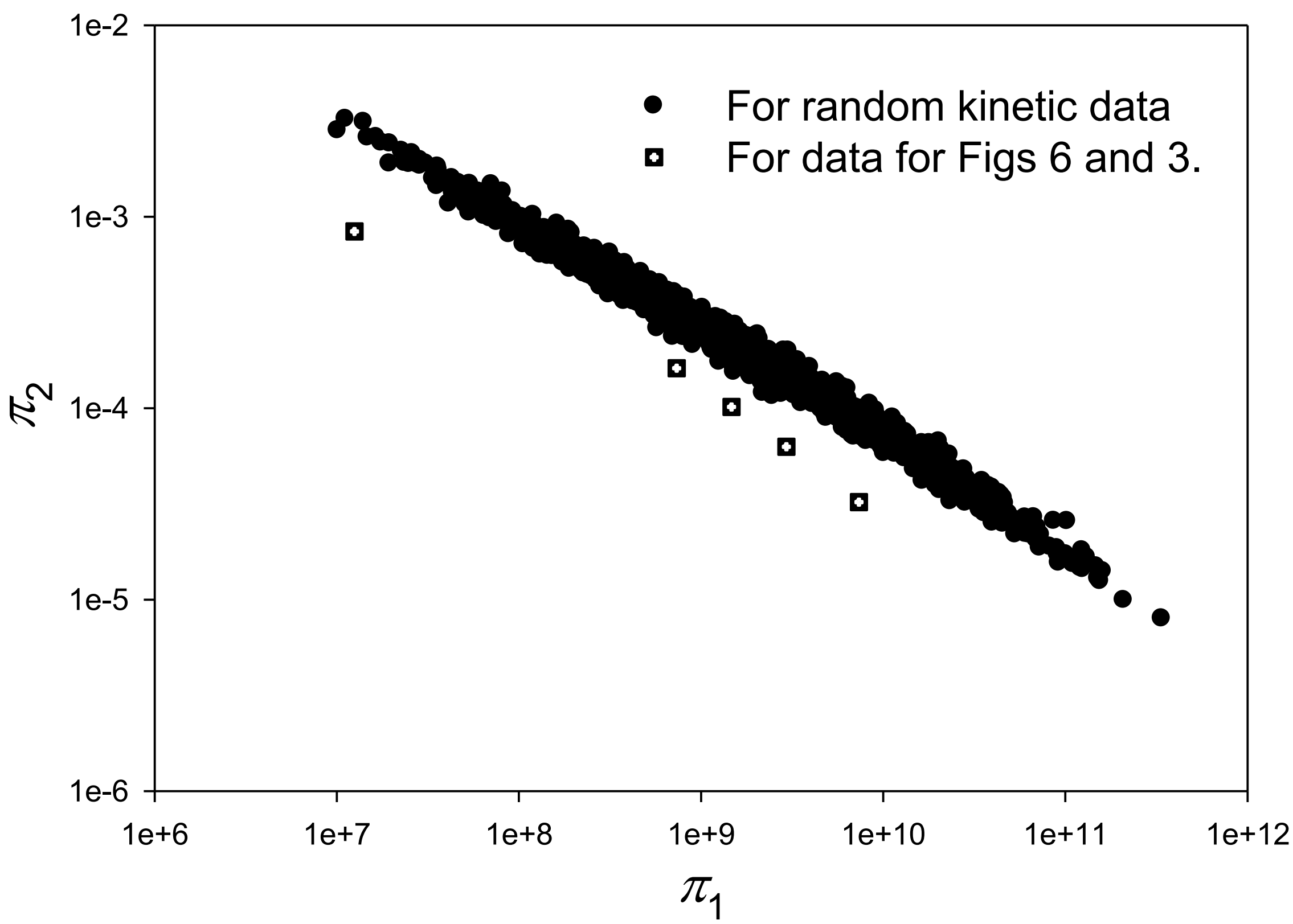

Fig. 8. An additional scaling law uncovered by the numerical formulation